\documentclass[%
 reprint,
nofootinbib,
 amsmath,amssymb,
 aps,
]{revtex4-2}
\usepackage{amsmath}
\usepackage[colorlinks=true, linkcolor=blue, citecolor=blue]{hyperref}
\usepackage{graphicx}
\usepackage{dcolumn}
\usepackage{bm}

\begin{document}

\preprint{APS/123-QED}

\title{Structured Illumination Scanning Thermography (SISTER)}

\author{Pengfei Zhu}
 \email{pengfei.zhu@bam.de}
\affiliation{%
 Bundesanstalt für Materialforschung and -prüfung (BAM), 12205 Berlin, Germany
}%

\author{Julien Lecompagnon}
\affiliation{%
	Bundesanstalt für Materialforschung and -prüfung (BAM), 12205 Berlin, Germany
}%

\author{Philipp Daniel Hirsch}
\affiliation{%
	Bundesanstalt für Materialforschung and -prüfung (BAM), 12205 Berlin, Germany
}%

\author{Mathias Ziegler}
\affiliation{%
	Bundesanstalt für Materialforschung and -prüfung (BAM), 12205 Berlin, Germany
}%

\date{\today}

\begin{abstract}
  Conventional non-invasive photothermal imaging techniques are fundamentally constrained by the diffusive nature of heat transport, which causes severe energy dissipation during subsurface reconstruction. Although modulation-based approaches partially mitigate this limitation by encoding depth information into phase delay and amplitude attenuation, they remain inherently restricted by repeated temporal excitation, long acquisition times, and stitching artifacts in large-area inspection.
  In this work, we propose a structured illumination scanning thermography (SISTER) framework that replaces conventional temporal modulation with continuous spatial scanning under static structured illumination. The key theoretical insight is that heat diffusion is governed by a Markov semigroup, while sample motion transforms static spatial illumination into an equivalent temporal excitation through a Galilean coordinate transformation. This formulation enables dynamic-to-static reconstruction without repeated temporal modulation and provides a unified interpretation of spatial scanning and conventional signal modulation.
  A scanning system is integrated to implement the proposed framework together with a dynamic-to-static reconstruction algorithm for continuous subsurface defect inspection. Both numerical simulations and experimental results demonstrate that the proposed method significantly improves spatial continuity, signal-to-noise ratio, and detection capability while effectively eliminating stitching artifacts and reducing acquisition complexity.
  The proposed SISTER framework establishes a unified theoretical foundation for scanning photothermal imaging and provides a practical paradigm for high-efficiency, large-scale industrial non-destructive testing.
\end{abstract}

\maketitle

\section{Introduction}
Non-invasive photothermal imaging has emerged as a powerful tool for probing the internal structure~\cite{1,2,3} and dynamics of materials~\cite{4,5} with high spatial and temporal resolution. By converting absorbed light into localized heat and monitoring the ensuing thermal waves with a non-scanning multi-array infrared (IR) camera, it provides a unique window into subsurface features without physically perturbing the sample~\cite{6,7,8}. This process is fundamentally governed by thermal diffusion-wave theory~\cite{9,10,11}, which describes the gradient-driven propagation of heat and its accumulation effect at discontinuous interfaces~\cite{12,13,14}. However, such characteristics inherently lead to the absence of a well-defined wavefront and strong dissipative losses during heat transport, limiting the photothermal imaging’s ability to resolve deeper structures~\cite{15,16} or containing higher spatial resolution~\cite{17,18} comparable to optical and acoustic imaging techniques~\cite{19,20,21}. 

The most used photothermal imaging technique in industrial applications is pulsed thermography~\cite{22,23}, which utilizes an external heat source with short pulse-width (e.g., pulse laser or flash lamps) to simulate the Dirac pulse. The single pulse excitation can generate transient heat gradient and then the thermal feature can be captured before the temperature attenuation~\cite{24,25}. In addition, the single pulse excitation has broadband spectrum, which offers different depth information based on the quantitative relationship between the frequency $\omega$ and the thermal diffusion length $\mu$, $\mu = \sqrt{2\alpha /\omega}$ (where $\alpha$ is the thermal diffusivity)~\cite{26,27}. However, owing to the parabolic nature of the heat diffusion equation, $\partial T / \partial t = \alpha \partial^2 T / \partial^2 x$, thermal perturbations spread rapidly, with their amplitude decaying as $T \propto \exp\left(-x^2/(4\alpha t)\right)$. As a result, thermal features originating from deeper layers become obscured by camera noise and environmental fluctuations~\cite{28,29}. In addition, the broadband temporal excitation simultaneously activates all spatial diffusion modes, whose strongly dissipative and non-orthogonal nature leads to an intrinsic superposition of thermal responses from different depths and length scales~\cite{30}. As a result, the transient temperature signal is fundamentally aliased and cannot be uniquely mapped to subsurface features~\cite{31}.

Signal modulation techniques can selectively excite thermal diffusion modes at a chosen frequency $\omega$, thereby controlling thermal wave propagation and enabling partial separation of contributions from different depths~\cite{32,33}. The first signal modulation technique in photothermal imaging is lock-in thermography (LIT)~\cite{34,35}, which employs a sinusoidal excitation, $Q=Q_0(1+\sin (\omega t))$, where the first term represents the DC component. This approach effectively mitigates signal aliasing and the low signal-to-noise ratio (SNR) inherent in pulsed thermography. However, LIT suffers from low imaging speed, requiring multiple modulation cycles to reach a steady-state signal. Moreover, a single LIT measurement provides information only from a specific depth, corresponding to the modulation frequency. Subsequently, thermal-wave radar (TWR) techniques~\cite{36,37} were proposed to improve both the dynamic range and depth resolution. More recently, chirp-pulsed TWR~\cite{38}, multi-frequency lock-in thermography~\cite{7}, and other frequency-sweep techniques~\cite{39} have been developed to extend detection depth and enhance defect contrast. However, these signal modulation techniques have largely remained confined to laboratory settings and are challenging to implement in industrial environments. The primary limitation is extremely long excitation period required for even a small area. In addition, complex sample geometries induce uneven heating and background noise, further degrading detection accuracy. In industrial practice, large samples are often divided into small sub-areas for measurement, which are subsequently stitched into a composite image. However, this approach introduces strong boundary artifacts. While stitching is straightforward for flat samples, it becomes considerably more challenging for complex geometries. Moreover, residual thermal imbalances between sub-areas persist, hindering further depth reconstruction.

\begin{figure*}[t]
	\centering
	\includegraphics[width=\textwidth]{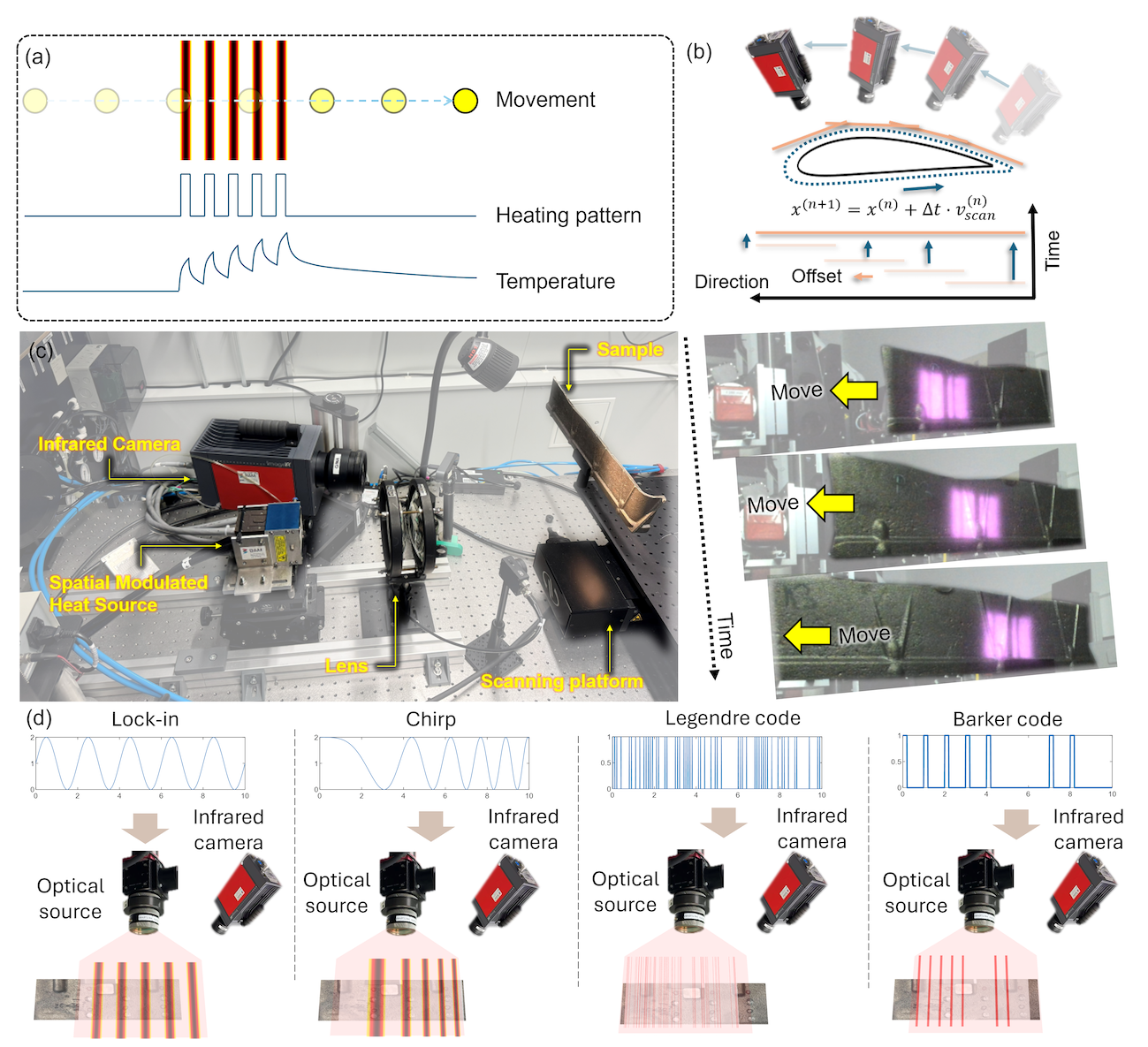}
	\caption{Structured illumination scanning thermography (SISTER) framework.
		(a) Principle of spatial-to-temporal modulation conversion. A static structured heating pattern is transformed into an effective temporal excitation through continuous sample motion, resulting in a periodic thermal response governed by heat diffusion.
		(b) Dynamic-to-static data reconstruction (DSDR). Thermal frames acquired during scanning are mapped back to the material coordinate system using the known scanning velocity, enabling recovery of the equivalent static structured-illumination response.
		(c) Experimental implementation of the SISTER system. An infrared camera records the thermographic response while the specimen is translated on a scanning platform beneath a spatially modulated heat source. Representative photographs illustrate continuous inspection during sample motion.
		(d) Comparison between conventional temporal modulation schemes and the proposed spatial coding strategy. Sinusoidal lock-in, chirped, Legendre-coded, and Barker-coded excitations are realized as stationary spatial illumination patterns that become equivalent temporal modulations under scanning, enabling frequency-selective thermographic measurements without repetitive temporal heating.}\label{fig1}
\end{figure*}

To address these challenges, we present structured illumination scanning thermography (SISTER), which converts conventional temporal modulation into spatial modulation via linear scanning (see Fig.~\ref{fig1}). By employing linear scanning, photothermal imaging is no longer restricted to fixed local inspection, but becomes a motion-flexible, signal-modulated imaging technique. In details, SISTER uses a spatial modulated heat source to generate structured illumination. Static structured illumination cannot generate the effect of temporal-to-spatial modulation. Therefore, a path-parameterized scanning platform should be involved. It carries the sample to move. The relevant parameter setup will be discussed in the following sections. Then, a dynamic-to-static data reconstruction (DSDR) algorithm is conducted to convert the linear scanning results to spatially coded thermography results. Finally, experimental and simulation results were analyzed to compare the proposed SISTER technique and conventional linear scanning thermography (LST) technique.
\section{Theory}\label{Section II}
The proposed structured illumination scanning thermography (SISTER) is fundamentally different from conventional signal-modulated thermography. Instead of generating temporal modulation electronically, SISTER converts a static spatial illumination into an equivalent temporal excitation through continuous sample motion. This section establishes the theoretical foundation of this equivalence from the perspectives of diffusion semigroup theory and coordinate transformation.
\subsection{Markov Semigroup-Driven Diffusion}
Heat conduction is governed by 
\begin{equation} 
	\frac{\partial T(\mathbf{x},t)}{\partial t} = \alpha\nabla^2T(\mathbf{x},t) + Q(\mathbf{x},t), 
	\label{eq:heat} 
\end{equation} 
where $T$ denotes the temperature field, $\alpha$ is the thermal diffusivity, and $Q$ represents the external heat source. Define the infinitesimal generator 
$
	\mathcal{A}=\alpha\nabla^2. 
$
Equation (\ref{eq:heat}) becomes 
\begin{equation} 
	\frac{dT}{dt}=\mathcal{A}T+Q. 
	\label{eq:operator} 
\end{equation} 

According to the Hille--Yosida theorem, the diffusion operator generates a strongly continuous Markov semigroup, $ P(t)=e^{t\mathcal{A}}.$ Consequently, the temperature evolution admits the mild solution 
\begin{equation} 
	T(t) = P(t)T_0 + \int_0^t P(t-\tau)Q(\tau)d\tau, 
	\label{eq:duhamel} 
\end{equation} 
where $T_0$ denotes the initial temperature distribution. Unlike wave propagation, the diffusion semigroup satisfies 
\begin{equation} 
	P(t+s)=P(t)P(s), 
	\label{eq:semigroup} 
\end{equation} 
which indicates that thermal evolution is first-order, memoryless, contractive, and completely determined by the current state. Therefore, every newly acquired thermal frame can be interpreted as one additional application of the same diffusion operator.

\subsection{Galilean Transformation of Structured Illumination}
Unlike conventional lock-in thermography, the projected structured illumination is spatially static. Temporal modulation arises solely from sample motion. Let $x$ denote the laboratory coordinate, and $X$ the material coordinate attached to the moving sample. For a constant scanning velocity $v$, 
\begin{equation} 
	x=X+vt. 
	\label{eq:galilean} 
\end{equation} 

Suppose the spatial modulation source projects a fixed spatial pattern $ Q(x)=P(x),$ where $P(x)$ denotes an arbitrary structured illumination. The heat source experienced by a material point becomes 
\begin{equation} 
	Q(X,t) = P(X+vt). 
	\label{eq:moving_source} 
\end{equation} 

Equation (\ref{eq:moving_source}) demonstrates that sample motion converts a purely spatial modulation into a temporal excitation without changing the optical source. Define the scanning operator 
\begin{equation} 
	(S_vP)(X,t) = P(X+vt). 
	\label{eq:scan} 
\end{equation} 
The diffusion equation therefore becomes 
\begin{equation} 
	\frac{dT}{dt} = \mathcal{A}T + S_vP. 
	\label{eq:scan_heat} 
\end{equation}

Since the Laplacian operator is translation invariant, 
\begin{equation} 
	\mathcal{A}S_v=S_v\mathcal{A}. 
	\label{eq:commute} 
\end{equation} 
Therefore, 
\begin{equation} 
	P(t)S_v = S_vP(t). 
	\label{eq:commute2} 
\end{equation} 

Equation (\ref{eq:commute2}) proves that performing spatial scanning before diffusion is mathematically equivalent to applying diffusion before spatial translation. Consequently, continuous scanning does not alter the intrinsic thermal transport process. Instead, it merely changes the coordinate representation of the heat source. Substituting Eq.~(\ref{eq:scan}) into Eq.~(\ref{eq:duhamel}) yields 
\begin{equation} 
	T(t) = P(t)T_0 + \int_0^t P(t-\tau) S_vP(\tau) d\tau, 
	\end{equation} 
which establishes the theoretical foundation of SISTER.

\subsection{Dynamic-to-Static Data Reconstruction (DSDR) and Spectral Interpretation}
The infrared camera measures the temperature in the laboratory coordinate, 
\begin{equation} 
	I(x,t) = T(x-vt,t). 
	\label{eq:image} 
\end{equation} 

Recovering the temperature evolution in the material frame requires only the inverse coordinate transformation, 
\begin{equation} 
	T_{\mathrm{rec}}(X,t) = I(X+vt,t). 
	\label{eq:inverse} 
\end{equation} 

Equation (\ref{eq:inverse}) forms the mathematical basis of the proposed dynamic-to-static reconstruction algorithm.
Expanding the structured illumination into its spatial Fourier series, 
\begin{equation} 
	P(x) = \sum_k P_k e^{ikx}, 
\end{equation} 
gives 
\begin{equation} 
	P(X+vt) = \sum_k P_k e^{ikX} e^{ikvt}. 
\end{equation} 

Hence every spatial Fourier mode naturally becomes a temporal harmonic, $\omega=vk.$ For each Fourier mode, the diffusion equation admits the steady-state response 
\begin{equation} 
	\Theta_k = \frac{P_k} {\alpha k^2+i\omega}, 
\end{equation} 
which becomes 
\begin{equation} 
	H(k,v) = \frac{1} {\alpha k^2+ikv} 
	\label{eq:H} 
\end{equation} 
after substituting $\omega=vk$. Equation (\ref{eq:H}) is identical to the transfer function of conventional temporal modulation, except that the temporal frequency is generated by spatial scanning instead of electronic modulation. Therefore, SISTER realizes frequency-selective thermal excitation without repeated temporal heating.
\begin{figure*}[t]
	\centering
	\includegraphics[width=\textwidth]{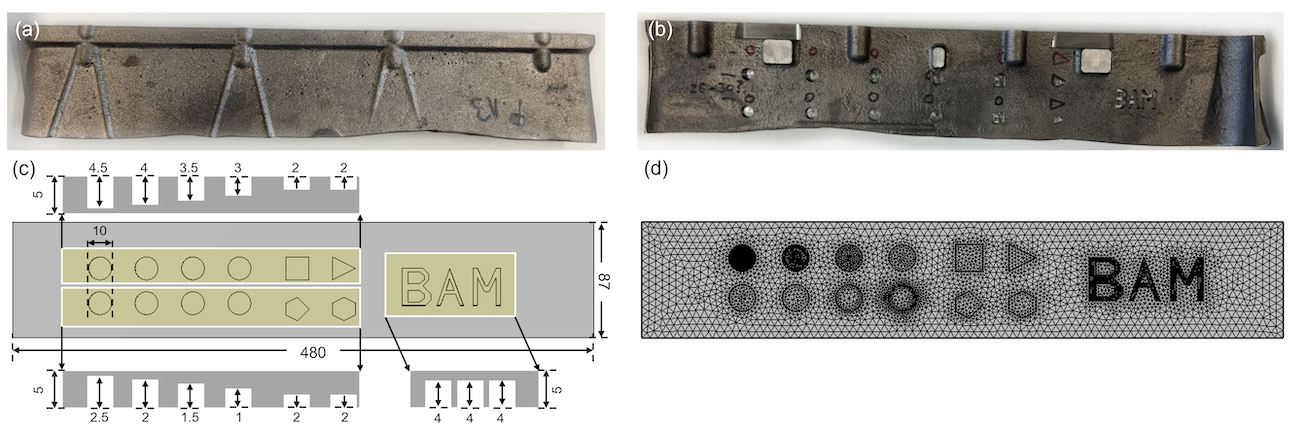}
	\caption{Test specimen and defect design used for experimental validation.
		(a) Front-side view of the cast-metal specimen extracted from an industrial casting component.
		(b) Back-side view showing the artificially machined defect patterns. Standard flat-bottom holes (FBHs) with different lateral geometries and depths, together with a defect pattern forming the “BAM” logo, were introduced to simulate representative subsurface flaws.
		(c) Schematic of the specimen geometry and defect distribution. The specimen has overall dimensions of 480 mm × 87 mm. Circular, square, triangular, pentagonal, and hexagonal FBHs were machined on the rear surface with various remaining wall thicknesses to evaluate defect detectability as a function of depth and shape. The “BAM” pattern consists of FBHs with a constant remaining wall thickness of 4 mm.
		(d) Finite-element mesh and numerical model employed for thermal simulations, reproducing the geometry and defect layout of the experimental specimen.}\label{fig2}
\end{figure*}
\section{Experimental and Simulation Setup}
\subsection{Experimental Setup}
Figure~\ref{fig1} illustrates the concept and experimental implementation of the proposed structured illumination scanning thermography (SISTER). Instead of generating temporal modulation through repetitive excitation, SISTER employs a static spatially structured heating pattern and converts it into an equivalent temporal excitation via continuous sample translation. As shown in Fig.~\ref{fig1}(a), a moving sample sequentially experiences different regions of the spatial heating pattern, resulting in a temporally modulated thermal response. Figure~\ref{fig1}(b) presents the dynamic-to-static data reconstruction (DSDR) process, in which infrared frames acquired in the laboratory coordinate system are mapped back to the material coordinate system according to the scanning velocity, thereby recovering the thermal evolution under stationary structured illumination. The experimental setup is shown in Fig.~\ref{fig1}(c), consisting of a spatially modulated optical heat source, an infrared camera, and a motorized scanning platform. Representative photographs demonstrate continuous inspection during sample translation. Figure~\ref{fig1}(d) further highlights the generality of the proposed framework by showing several representative coding schemes, including lock-in, chirp, Legendre, and Barker excitations. Within the SISTER framework, these conventional temporal modulation signals are implemented as spatial illumination patterns, establishing a unified interpretation of signal-modulated thermography based on spatial coding and scanning motion.

The infrared camera was from Infratec ImageIR 9300 equipped with a cooled indium antimonide (InSb) focal-plane array (1280 $\times$ 1024 pixels, mid-wave infrared, 3-5 $\mu$m), providing a noise-equivalent temperature difference below 30 mK at 30 $^\circ$C. The structured illumination source was realized using a one-dimensional VCSEL array (ULM Photonics, ULM850-14-TT-N0112U) operating at a central wavelength of 850~nm. The array consists of 12 emitters arranged linearly with common cathodes, producing a spatially distributed heating pattern. The emitted optical power was modulated through the spatial arrangement of the VCSEL elements rather than by temporal driving signals. Owing to the approximately 850~nm emission wavelength, narrow spectral bandwidth ($<1$~nm), and beam divergence of $20^\circ$--$30^\circ$, the VCSEL array provides stable and well-defined structured illumination suitable for scanning photothermal excitation. During inspection, the projected spatial pattern remained stationary while the sample was translated by a motorized scanning stage, thereby converting the spatial coding into an equivalent temporal excitation.

The investigated specimen was a cast metallic component provided within the BAM project TT-Guss (“Non-destructive testing of cast products by laser thermography”). The sample contains representative casting defects embedded beneath the surface, including defect structures typically encountered in industrial castings, such as porosity, shrinkage-related discontinuities, and internal inhomogeneities. These subsurface defects locally alter the heat diffusion process and therefore provide a suitable benchmark for evaluating the capability of the proposed structured illumination scanning thermography (SISTER) method. The specimen was selected to represent realistic industrial inspection scenarios, where reliable detection of buried defects within large-area cast components remains a critical challenge for quality assurance. In this work, a section of the original casting component was extracted and used as the inspection specimen. Flat-bottom holes (FBHs) with calibrated geometries were machined on the rear surface to mimic subsurface defects. The artificial defects were designed with different diameters, depths, and shapes, including a defect pattern forming the “BAM” logo, providing a representative benchmark for evaluating the defect detectability and spatial resolution of the proposed method (see Fig.~\ref{fig2}).
\begin{figure*}[t]
	\centering
	\includegraphics[width=\textwidth]{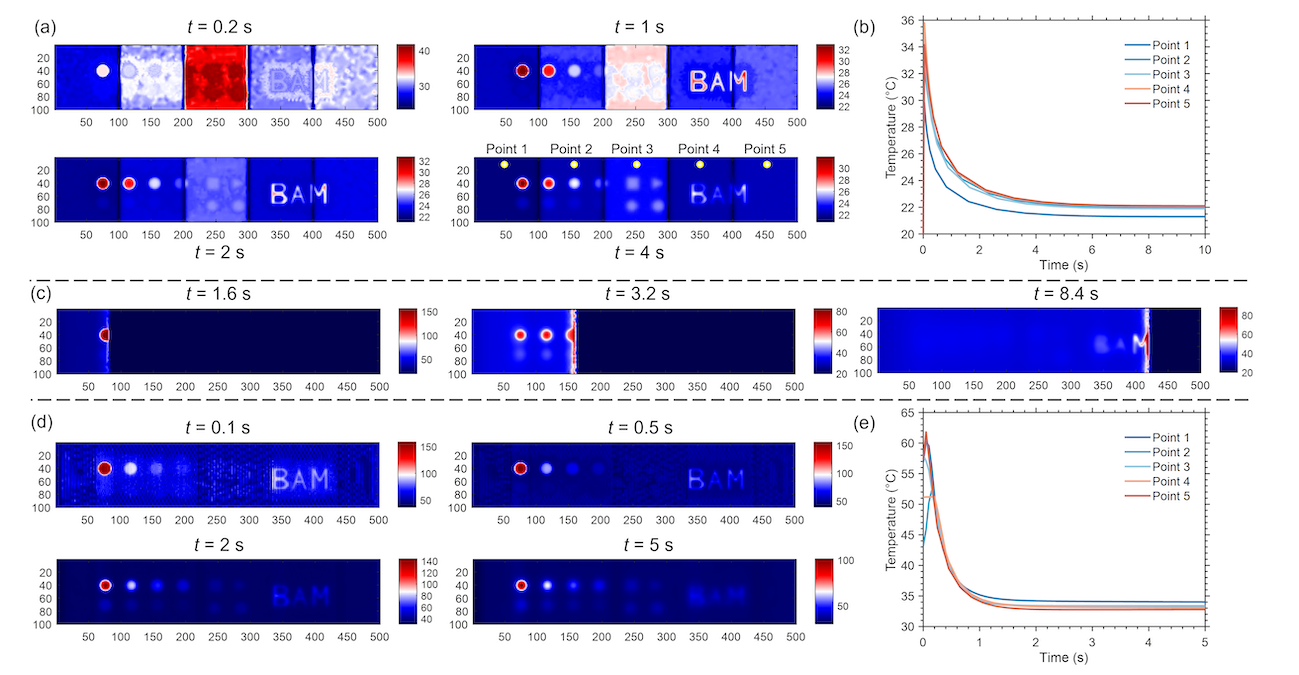}
	\caption{Comparison between pulsed thermography and linear scanning thermography (LST) with dynamic-to-static reconstruction.
		(a) Stitched pulsed thermography images acquired at different times after excitation, showing the temporal evolution of the thermal contrast associated with the artificial defects.
		(b) Temperature decay curves extracted from five non-defective locations indicated in (a). Although all points exhibit similar cooling behavior, noticeable temperature variations remain due to spatially non-uniform heating and image stitching artifacts.
		(c) Raw LST results at representative time instants during continuous specimen translation. The thermal response is strongly localized around the moving heat source and does not directly reveal the complete defect distribution.
		(d) Reconstructed LST images obtained using the dynamic-to-static data reconstruction (DSDR) algorithm. The transformation from the laboratory coordinate system to the material coordinate system enables recovery of the stationary thermal response, revealing the complete defect pattern over the entire inspection area.
		(e) Temperature histories obtained from the same five locations after DSDR reconstruction.}\label{fig3}
\end{figure*}
\subsection{Simulation Setup}
To validate the proposed structured illumination scanning thermography (SISTER) framework, finite-element simulations were performed using the transient heat conduction equation

\begin{equation}
	\rho c_p \frac{\partial T}{\partial t}
	=
	\nabla \cdot
	\left(
	k \nabla T
	\right)
	+
	Q(x,y,t),
\end{equation}
where the thermal conductivity, density, and specific heat capacity were set to
$k=16.2~\mathrm{W/(m\cdot K)}$,
$\rho=7900~\mathrm{kg/m^3}$,
and
$c_p=477~\mathrm{J/(kg\cdot K)}$,
respectively.

A moving structured heat source was applied on the front surface of the specimen. The total heat flux distribution was defined as

\begin{equation}
	Q(x,y,t)
	=
	a_{1}(x,y,t)
	+a_{2}(x,y,t)
	+a_{3}(x,y,t)
	+a_{4}(x,y,t),
\end{equation}
where
\begin{equation}
	a_{1}
	=
	10^{6}
	\exp
	\left[
	-\frac{(x-vt)^2}{2\sigma^2}
	\right],
\end{equation}

\begin{equation}
	a_{2}
	=
	10^{6}
	\exp
	\left[
	-\frac{(x+5v-vt)^2}{2\sigma^2}
	\right],
\end{equation}

\begin{equation}
	a_{3}
	=
	10^{6}
	\exp
	\left[
	-\frac{(x+8v-vt)^2}{2\sigma^2}
	\right],
\end{equation}

\begin{equation}
	a_{4}
	=
	10^{6}
	\exp
	\left[
	-\frac{(x+10v-vt)^2}{2\sigma^2}
	\right].
\end{equation}

The scanning velocity and Gaussian beam width were chosen as
$v=10~\mathrm{mm/s}$
and
$\sigma=1.75~\mathrm{mm}$,
respectively. All remaining boundaries were subjected to natural convective heat transfer with a heat transfer coefficient of 
$
	h=10~\mathrm{W/(m^2\cdot K)}.
$
The transient temperature field was computed over the time interval

\begin{equation}
	t =
	\mathrm{range}
	\left(
	0,
	\frac{1}{v},
	\frac{500+15v}{v}
	\right),
\end{equation}
which corresponds to the complete passage of the structured illumination pattern across the specimen. For linear scanning thermography simulation, only $a_1$ was used as the heat source. For pulsed thermography, the duration of heat source simulation is 0.02 s with a power of $10^6~\mathrm{W/m^2}$. The sample was divided equally into 5 parts with a width of 100 mm.
\subsection{Parameter Design for SISTER}
The parameters of the structured illumination pattern were selected according to the equivalence between conventional temporal modulation and the proposed SISTER framework (see Table~\ref{tab:sister_parameter}). In the temporal domain, a modulation signal is characterized by the observation time $t_{\mathrm{obs}}$, pulse width $\Delta t$, and camera sampling frequency $f_c$. Under uniform scanning motion, these quantities are directly mapped into spatial parameters through
\begin{equation}
	W=v\,t_{\mathrm{obs}},
\end{equation}
and
\begin{equation}
	\Delta W=v\,\Delta t,
\end{equation}
where $W$ denotes the total length of the spatial heating pattern and $\Delta W$ represents the width of an individual heating element. Consequently, the temporal excitation sequence can be interpreted as a stationary spatial code observed by a moving specimen.

For accurate dynamic-to-static reconstruction, the scanning velocity must be selected to satisfy both the observation-length requirement and the spatial sampling constraint. Assuming a camera frame rate $f_c$ and image spatial resolution $\Delta x$, the maximum allowable velocity is approximately

\begin{equation}
	v \le
	\min
	\left(
	\frac{\mathrm{FOV}}{t_{\mathrm{obs}}},
	\Delta x\, f_c
	\right),
\end{equation}
which ensures that each spatial location is sampled by at least one thermal frame and prevents motion-induced undersampling. Correspondingly, the effective spatial resolution of SISTER can be expressed as

\begin{equation}
	\Delta x_{\mathrm{eff}}
	=
	\max
	\left(
	\Delta x,
	\frac{v}{f_c}
	\right),
\end{equation}
indicating that both the imaging resolution and the scanning speed jointly determine the recoverable spatial information.
\begin{table}[t]
	\caption{Parameter correspondence between conventional temporal modulation (TM) and the proposed SISTER framework.}
	\label{tab:sister_parameter}
	\centering
	\begin{tabular}{lll}
		\hline
		Parameter & TM & SISTER \\
		\hline
		Observation time & $t_{\mathrm{obs}}$ &
		$W=v\,t_{\mathrm{obs}}$ \\
		
		Pulse width &
		$\Delta t$ &
		$\Delta W=v\,\Delta t$ \\
		
		Modulation period &
		$T$ &
		$\Lambda=v\,T$ \\
		
		Modulation frequency &
		$f=1/T$ &
		$f=v/\Lambda$ \\
		
		Camera sampling interval &
		$1/f_c$ &
		$\max(1/f_c,\Delta x/v)$ \\
		
		Spatial resolution &
		$\Delta x$ &
		$\max\!\left(\Delta x,\frac{v}{f_c}\right)$ \\
		
		Scanning velocity &
		--- &
		$v=W/t_{\mathrm{obs}}$ \\
		
		Pattern length &
		--- &
		$W=v\,t_{\mathrm{obs}}$ \\
		
		Pattern element width &
		--- &
		$\Delta W=v\,\Delta t$ \\
		\hline
	\end{tabular}
\end{table}
In the present study, the scanning velocity was fixed at
$v=10~\mathrm{mm/s}$,
which provides sufficient temporal sampling during specimen translation while maintaining practical inspection efficiency. The Gaussian width was chosen as
$\sigma=1.75~\mathrm{mm}$,
resulting in partially overlapping heating spots and a smooth spatial energy distribution. Four Gaussian heating components were used to emulate a spatially coded excitation sequence. The spacing between adjacent Gaussian peaks determines the equivalent modulation period, while the Gaussian width controls the duty cycle and thermal overlap between neighboring excitation elements. This configuration was found to provide a favorable compromise between thermal signal strength, spatial localization, and reconstruction stability.
\begin{figure*}[t]
	\centering
	\includegraphics[width=\textwidth]{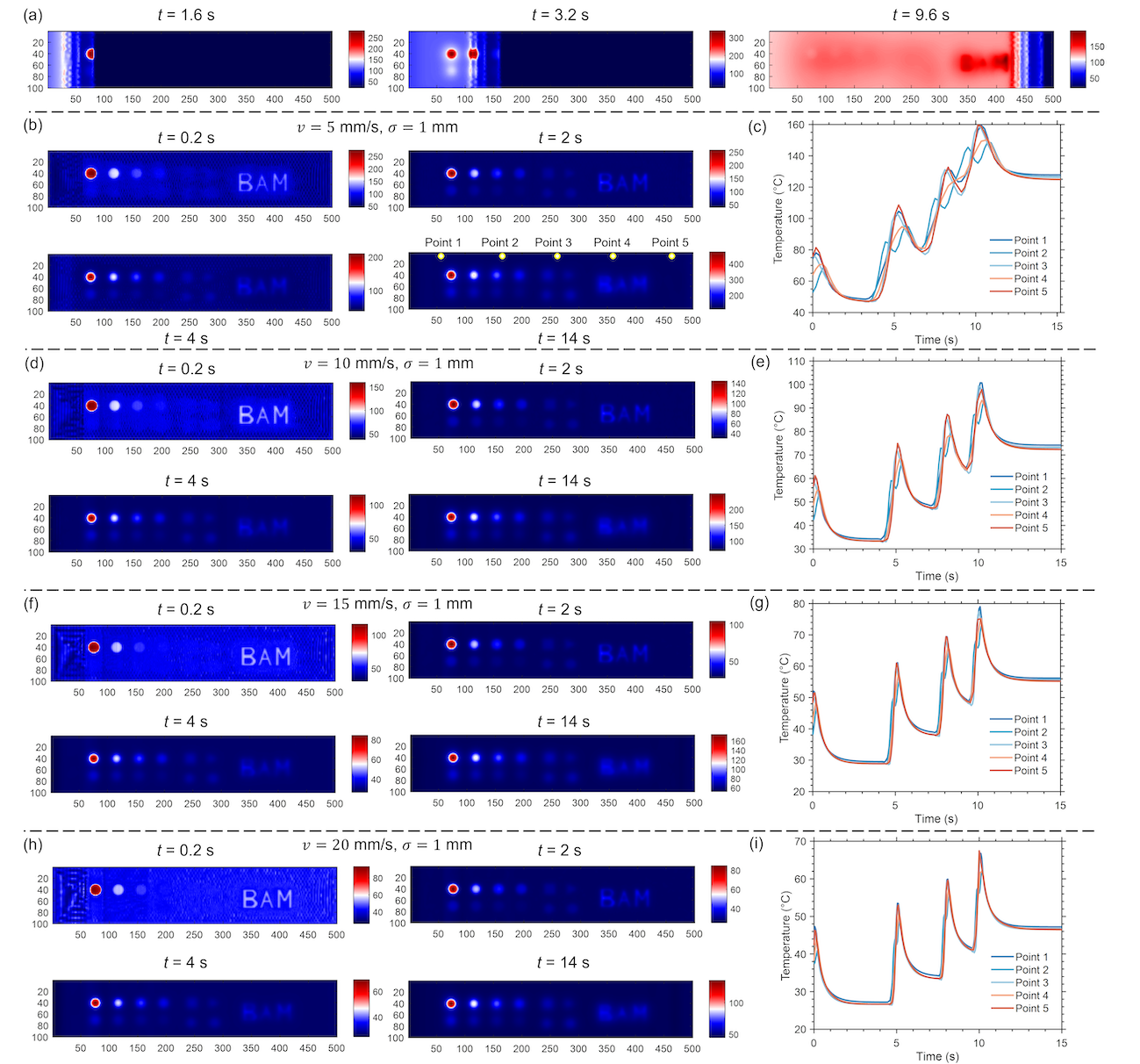}
	\caption{Influence of scanning velocity on SISTER reconstruction performance.
		(a) Representative raw structured illumination scanning thermography (SISTER) data acquired in simulation.
		(b), (d), (f), and (h) DSDR-reconstructed SISTER images obtained for scanning velocities of $v=5,, 10, 15$, and $20$ mm/s, respectively, with $\sigma=1~\mathrm{mm}$.
		(c), (e), (g), and (i) Temperature histories extracted from the five locations indicated in the corresponding reconstructed images. The multiple thermal peaks correspond to the sequential passage of the structured heating elements.}\label{fig4}
\end{figure*}
\section{Results and Discussion}
\subsection{Simulation Results}
Figure~\ref{fig3} compares conventional pulsed thermography (PT) and the proposed linear scanning thermography combined with dynamic-to-static data reconstruction (LST+DSDR). As shown in Fig.~\ref{fig3}(a), large-area PT inspection requires stitching multiple measurement regions, which introduces visible discontinuities and thermal imbalance between adjacent sub-images. These stitching artifacts persist throughout the cooling process and complicate subsequent quantitative analysis. This effect is further confirmed by the temperature histories extracted from five defect-free locations (Fig.~\ref{fig3}(b)), where noticeable differences are observed despite the points being located in nominally identical regions. The discrepancies originate primarily from non-uniform heating and accumulated stitching errors rather than intrinsic material variations.

In contrast, the raw LST measurements (Fig.~\ref{fig3}(c)) only capture localized thermal responses around the moving heat source and therefore cannot directly provide a global representation of the specimen. After applying the proposed DSDR algorithm, however, the continuously acquired scanning data are transformed into the material coordinate system, yielding stationary thermographic images (Fig.~\ref{fig3}(d)) that closely resemble the temporal evolution observed in conventional PT. More importantly, the reconstructed images exhibit excellent spatial continuity without any visible stitching boundaries. The temperature curves extracted from the same five locations (Fig.~\ref{fig3}(e)) show a highly consistent cooling behavior, demonstrating that the DSDR procedure effectively compensates for motion-induced distortions and restores a unified thermal field. These results indicate that LST+DSDR can reproduce the characteristic transient response of pulsed thermography while eliminating stitching artifacts and enabling continuous large-area inspection.
\begin{table*}[t]
	\caption{\label{tab:cnr_velocity}
		Contrast-to-noise ratio (CNR) obtained at different scanning velocities and defect depths.}
	\centering
	\begin{ruledtabular}
		\begin{tabular}{c c c c c c c c c}
			$v$ & \multicolumn{8}{c}{Defect depth} \\
			\cline{2-9}
			(mm/s) & 0.5 mm & 1.0 mm & 1.5 mm & 2.0 mm & 2.5 mm & 3.0 mm & 3.5 mm & 4.0 mm \\
			\hline
			5  & \textbf{6.46} & \textbf{2.93} & \textbf{1.67} & \textbf{0.99} & 0.48 & 0.23 & 0.07 & \textbf{0.05} \\
			10 & 6.25 & 2.83 & 1.60 & 0.97 & \textbf{0.50} & \textbf{0.25} & 0.09 & 0.02 \\
			15 & 6.14 & 2.73 & 1.57 & 0.95 & \textbf{0.50} & 0.24 & \textbf{0.10} & 0.01 \\
			20 & 6.08 & 2.67 & 1.53 & 0.94 & \textbf{0.50} & 0.23 & \textbf{0.10} & 0.01 \\
		\end{tabular}
	\end{ruledtabular}
\end{table*}

Figure~\ref{fig4} presents numerical simulations of the proposed SISTER framework for different scanning velocities. The raw SISTER data shown in Fig.~\ref{fig4}(a) are represented in the laboratory coordinate system, where the thermal response is confined to the vicinity of the moving structured heat source and the spatial distribution of subsurface defects cannot be directly interpreted. After applying the proposed DSDR algorithm, the thermal fields are transformed into the material coordinate system, yielding the reconstructed results shown in Figs.~\ref{fig4}(b), ~\ref{fig4}(d), ~\ref{fig4}(f), and ~\ref{fig4}(h). The reconstructed images clearly recover the complete defect distribution, including the flat-bottom holes and the “BAM” pattern, demonstrating the validity of the spatial-to-temporal modulation equivalence established in Section \ref{Section II}.

The simulations further reveal the influence of the scanning velocity on the reconstructed thermal response. As the scanning velocity increases from 5 to 20 mm/s, the interaction time between the structured illumination and the specimen decreases, leading to a systematic reduction in peak temperature and defect contrast. Nevertheless, the principal defect features remain identifiable over the entire investigated velocity range, indicating that the reconstruction process remains stable under accelerated scanning conditions.
The temperature histories extracted from five representative locations (Figs.~\ref{fig4}(c), ~\ref{fig4}(e), ~\ref{fig4}(g), and ~\ref{fig4}(i)) provide additional insight into the reconstruction mechanism. The multiple temperature peaks originate from the sequential passage of the individual Gaussian heating elements that constitute the structured illumination pattern. Increasing the scanning velocity compresses the temporal separation between adjacent peaks and decreases their amplitudes owing to the shorter local heating duration. Importantly, the reconstructed signals exhibit nearly identical temporal profiles at different locations, confirming that the DSDR procedure correctly compensates for the translational motion and reconstructs a spatially consistent thermal field. These numerical results verify that SISTER can reproduce the characteristic transient responses associated with structured thermal excitation while enabling continuous scanning-based inspection.

To quantitatively compare the defect contrast between different scanning speeds, the contrast-to-noise ratio (CNR) was employed, which is a commonly used index in NDT fields. The mathematical expression of CNR can be written as~\cite{40，41}:
\begin{equation}
	\mathrm{CNR} = |\frac{\mu_d - \mu_s}{\sigma_s}|,
\end{equation}
where $\mu_d$ and $\mu_s$ denote the mean values of the defect and sound regions, respectively, and $\sigma_s$ is the standard deviation of the sound region.

Table~\ref{tab:cnr_velocity} summarizes the quantitative evaluation of the reconstructed SISTER images at different scanning velocities. A clear trend can be observed: the contrast-to-noise ratio (CNR) decreases monotonically as the scanning velocity increases. This behavior is expected because a higher scanning velocity reduces the local interaction time between the structured illumination and the specimen, thereby decreasing the deposited thermal energy and the resulting defect contrast. Consequently, the signal amplitude becomes weaker and the reconstructed defects are progressively less distinguishable from the background. Despite this reduction in CNR, all investigated defect features remain detectable within the examined velocity range, confirming the robustness of the proposed reconstruction framework.
Interestingly, the temporal responses shown in Fig.~\ref{fig4} reveal an opposite tendency from the perspective of signal coding. As the scanning velocity increases, the reconstructed temperature histories increasingly resemble the designed chirp-pulse excitation profile. At low scanning velocities, thermal diffusion causes significant overlap between adjacent heating events, leading to broadened peaks and stronger inter-pulse coupling. In contrast, higher scanning velocities shorten the thermal interaction time and reduce the overlap between neighboring Gaussian heating elements, resulting in a sequence of temperature peaks that more faithfully reproduces the intended chirped excitation structure. Therefore, although increasing the scanning velocity inevitably sacrifices thermal contrast and CNR, it simultaneously improves the fidelity of the reconstructed modulation waveform. This trade-off highlights the dual role of scanning velocity in SISTER, acting not only as a parameter controlling inspection speed but also as a factor governing the balance between thermal signal strength and excitation-code fidelity.
\begin{table*}[t]
	\caption{\label{tab:cnr_sigma}
		Contrast-to-noise ratio (CNR) obtained with different Gaussian widths at a fixed scanning velocity of $v=10~\mathrm{mm/s}$.}
	\centering
	\begin{ruledtabular}
		\begin{tabular}{c c c c c c c c c}
			$\sigma$ & \multicolumn{8}{c}{Defect depth} \\
			\cline{2-9}
			(mm) & 0.5 mm & 1.0 mm & 1.5 mm & 2.0 mm & 2.5 mm & 3.0 mm & 3.5 mm & 4.0 mm \\
			\hline
			0.10 & 4.25 & 1.59 & 1.05 & 0.57 & 0.25 & 0.01 & \textbf{0.32} & \textbf{0.07} \\
			0.25 & 6.23 & 2.71 & 1.53 & 0.91 & \textbf{0.54} & 0.23 & 0.06 & 0.06 \\
			0.50 & 6.24 & 2.78 & 1.57 & 0.95 & \textbf{0.54} & 0.25 & \textbf{0.09} & 0.02 \\
			0.75 & 6.23 & 2.81 & 1.59 & 0.96 & 0.51 & 0.25 & \textbf{0.09} & 0.02 \\
			1.00 & 6.25 & 2.83 & 1.60 & 0.97 & 0.50 & 0.25 & \textbf{0.09} & 0.02 \\
			2.00 & \textbf{6.41} & \textbf{2.94} & \textbf{1.66} & \textbf{1.00} & 0.50 & \textbf{0.27} & \textbf{0.09} & 0.03 \\
		\end{tabular}
	\end{ruledtabular}
\end{table*}

Table~\ref{tab:cnr_sigma} summarizes the defect contrast-to-noise ratio (CNR) obtained for different Gaussian widths $\sigma$ at a fixed scanning velocity of 10 mm/s. For all investigated values of $\sigma$, the CNR decreases rapidly with increasing defect depth, reflecting the progressive attenuation of thermal diffusion waves and the reduced thermal contrast associated with deeper subsurface defects. The shallow defects (0.5–2.0 mm) exhibit the highest CNR values, whereas the detectability of defects deeper than 3 mm becomes increasingly challenging. Increasing the Gaussian width generally improves the thermal signal strength and leads to higher CNR values for most defect depths. In particular, the maximum CNR values for the shallow and intermediate defects are obtained at $\sigma = 2.0~\mathrm{mm}$, indicating that a broader heating profile promotes greater thermal energy deposition and improves defect visibility. However, the improvement becomes marginal for deeper defects, suggesting that the benefit of increasing $\sigma$ is ultimately limited by diffusion-induced attenuation.
From a signal-coding perspective, the Gaussian width also influences the fidelity of the reconstructed excitation waveform. A small $\sigma$ produces narrow and well-separated thermal peaks that more closely resemble the designed chirp-pulse sequence, whereas a larger $\sigma$ increases the overlap between adjacent heating events and smooths the temporal response. Therefore, the selection of $\sigma$ involves a trade-off between thermal contrast and waveform fidelity. While larger Gaussian widths provide improved defect detectability through higher CNR values, smaller widths yield reconstructed temperature histories that better preserve the temporal structure of the intended chirped excitation. Consequently, an intermediate Gaussian width offers a practical compromise, maintaining sufficient defect contrast while retaining the characteristic features of the chirp-coded SISTER excitation.
\begin{figure*}[t]
	\centering
	\includegraphics[width=\textwidth]{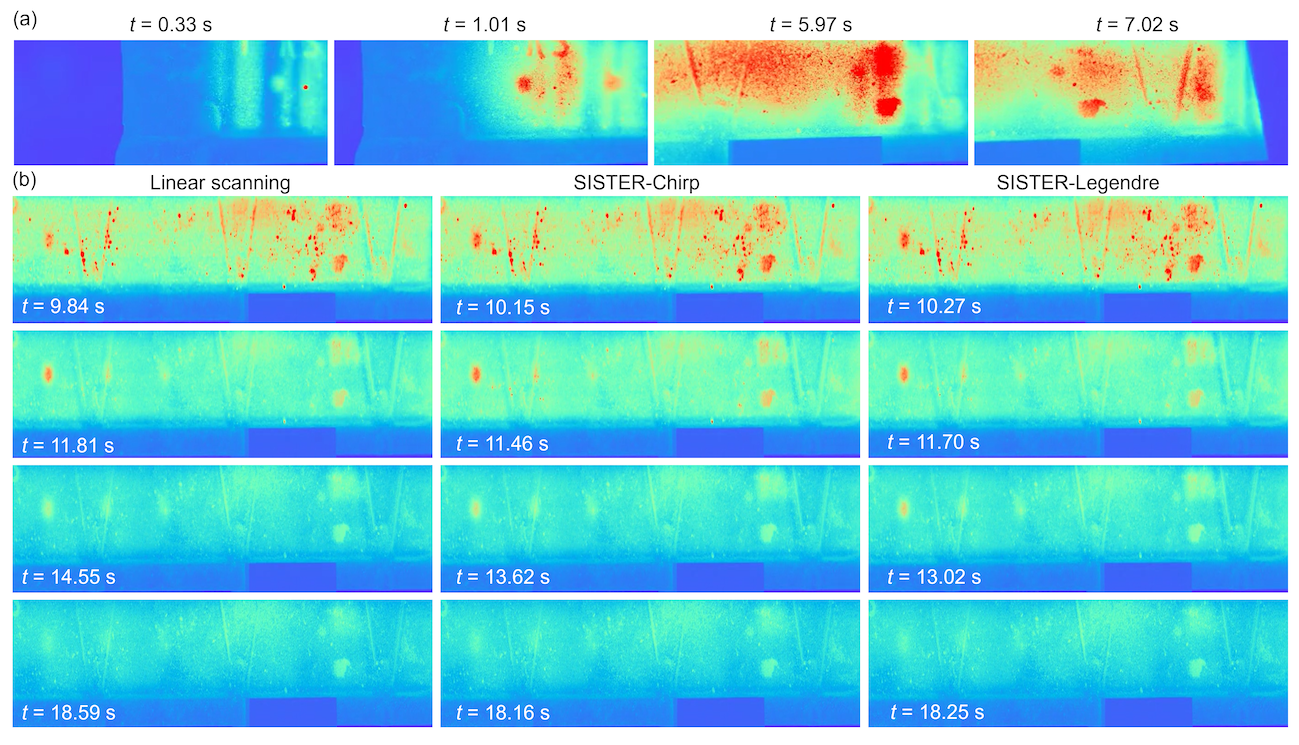}
	\caption{Experimental comparison of different spatial coding strategies in SISTER.
		(a) Representative raw SISTER measurements acquired during continuous scanning at different time instants, showing the thermal response in the laboratory coordinate system before reconstruction.
		(b) DSDR-reconstructed thermograms obtained using three VCSEL coding patterns: linear scanning (000001000000), SISTER-Chirp (101001000100), and SISTER-Legendre (111010111010). Representative reconstruction times are shown to illustrate the evolution of the recovered thermal field in the material coordinate system. The coded SISTER schemes provide enhanced thermal persistence and defect visibility compared with conventional linear scanning.}\label{fig5}
\end{figure*}
\subsection{Experimental Results}
Figure~\ref{fig5} compares conventional linear scanning thermography and two coded implementations of SISTER using chirp and Legendre illumination sequences. The raw SISTER measurements shown in Fig.~\ref{fig5}(a) are acquired in the laboratory coordinate system, where the thermal response evolves together with the scanning motion and the subsurface features cannot be directly interpreted. After applying the DSDR algorithm, all three excitation schemes successfully recover the thermal field in the material coordinate system, as shown in Fig.~\ref{fig5}(b). The reconstructed results demonstrate that the proposed framework is independent of the specific coding sequence and can accommodate arbitrary spatial excitation patterns.
A clear difference can nevertheless be observed in the temporal evolution of the reconstructed thermograms. For conventional linear scanning, corresponding to a single active VCSEL element (000001000000), the thermal signal rapidly decays after the heat source passes the inspection region. In contrast, the chirp-coded (101001000100) and Legendre-coded (111010111010) configurations generate multiple heating events distributed along the scanning direction, resulting in a prolonged thermal response and enhanced energy deposition. Consequently, defect indications remain visible for a longer time during the cooling process, particularly for weak-contrast regions. The results demonstrate that spatial coding can effectively control the temporal characteristics of the reconstructed thermal signal without modifying the scanning procedure itself.
More importantly, Fig.~\ref{fig5} confirms the generality of the SISTER concept. Through the combination of structured illumination and DSDR reconstruction, arbitrary temporal excitation schemes can be translated into equivalent spatial coding patterns. Therefore, conventional pulsed, chirped, coded, or multiplexed thermal excitations can be realized within a unified scanning framework, providing additional flexibility for improving defect detectability and optimizing thermographic inspection performance.
\begin{figure}[!t]
	\centerline{\includegraphics[width=\columnwidth]{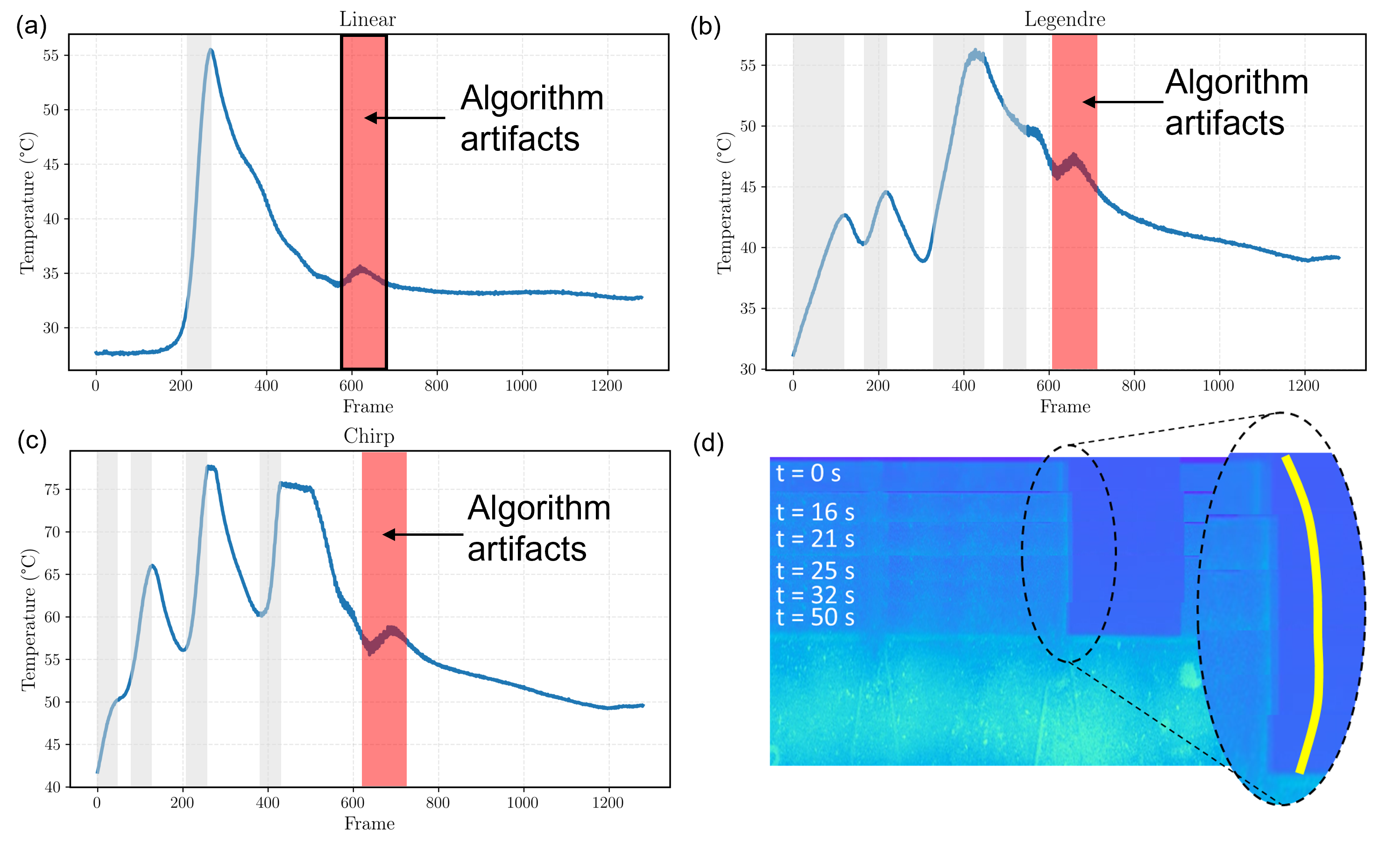}}
	\caption{Reconstruction artifacts induced by velocity mismatch in DSDR.
		(a)–(c) Temperature histories extracted from the DSDR-reconstructed datasets obtained with linear scanning, SISTER-Legendre, and SISTER-Chirp excitation patterns, respectively. The shaded gray regions indicate heating events associated with the structured illumination sequence. The red-highlighted regions correspond to reconstruction artifacts arising from imperfect velocity compensation.
		(d) Illustration of the origin of the reconstruction artifacts. A mismatch between the actual scanning-stage velocity and the velocity assumed in the DSDR algorithm introduces residual sample motion in the reconstructed coordinate system, leading to artificial temperature fluctuations and distortions in the reconstructed signals.}
	\label{fig6}
\end{figure}
Figure~\ref{fig6} investigates the influence of reconstruction-velocity mismatch on the performance of the DSDR algorithm. Figures~\ref{fig6}(a)–~\ref{fig6}(c) show the temperature histories reconstructed from the linear, Legendre-coded, and chirp-coded SISTER measurements, respectively. In all cases, the expected thermal responses associated with the illumination sequence are clearly observed in the early stage of the signal. However, an additional temperature fluctuation appears in the highlighted red region, which cannot be attributed to either heat diffusion or the designed excitation waveform. The artifact is consistently observed for all coding schemes, indicating that its origin is independent of the specific spatial encoding pattern.
The source of this behavior is illustrated in Fig.~\ref{fig6}(d). In practical measurements, a small discrepancy inevitably exists between the actual scanning-stage velocity and the velocity assumed in the DSDR reconstruction process. As a consequence, the transformed data are not mapped perfectly into the material coordinate system, leaving a residual translational motion in the reconstructed image sequence. This residual motion causes stationary structural features to drift slowly through the reconstructed field of view, producing artificial temperature variations that appear as spurious peaks or distortions in the reconstructed thermal histories. The effect becomes particularly noticeable at later reconstruction times, when the true thermal signal has substantially decayed and the reconstruction becomes increasingly sensitive to positional errors. These results indicate that accurate velocity calibration is critical for artifact-free DSDR reconstruction and highlight the importance of synchronization between the scanning system and the reconstruction algorithm in practical SISTER implementations.
\begin{figure*}[t]
	\centering
	\includegraphics[width=\textwidth]{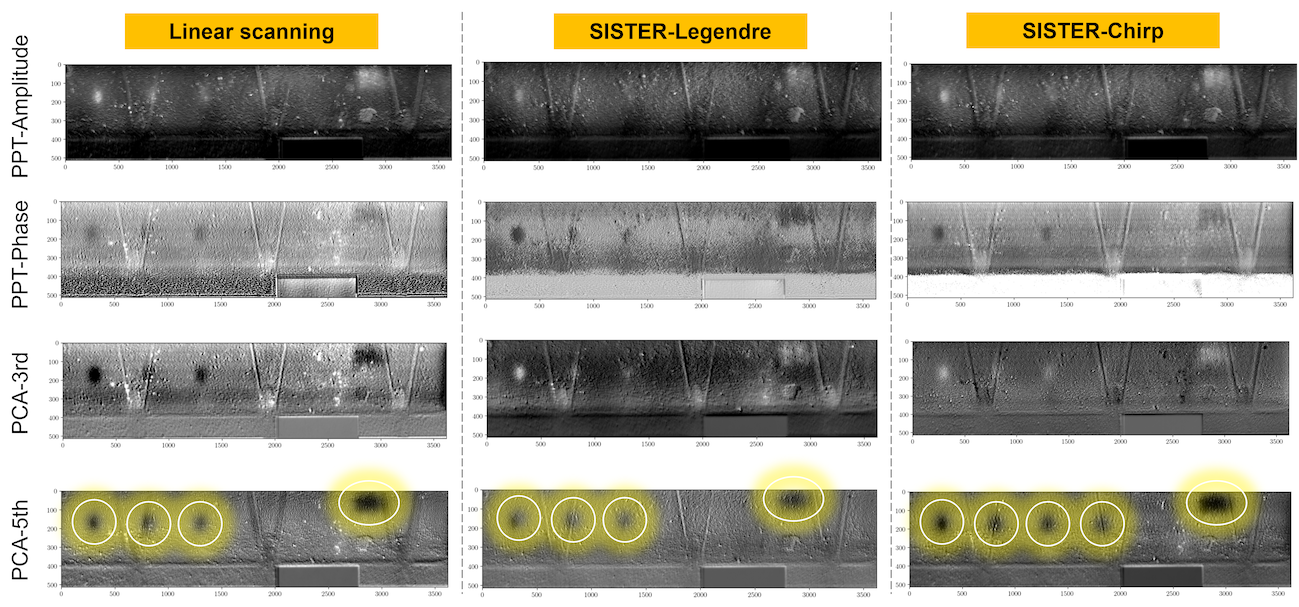}
	\caption{Comparison of image-processing methods applied to DSDR-reconstructed thermograms.
		DSDR-reconstructed datasets obtained using linear scanning, SISTER-Legendre, and SISTER-Chirp excitation patterns were processed by pulsed phase thermography (PPT) and principal component analysis (PCA). Rows correspond to PPT amplitude, PPT phase, the third principal component (PCA-3rd), and the fifth principal component (PCA-5th), respectively. The highlighted regions indicate representative subsurface features enhanced by the different excitation and post-processing schemes. The coded SISTER approaches provide improved defect visibility compared with conventional linear scanning, particularly in the PCA-based reconstructions.}\label{fig7}
\end{figure*}
Figure~\ref{fig7} compares the performance of different excitation schemes after DSDR reconstruction and subsequent thermographic post-processing including pulsed phase thermography (PPT)~\cite{22} and principal component analysis (PCA)~\cite{42}. For the PPT amplitude and phase images, all three excitation strategies reveal the dominant defect features with comparable contrast, although the coded SISTER approaches exhibit a more uniform thermal response over the inspected area. Greater differences emerge in the PCA results. The third principal component (PCA-3rd) enhances several major defect indications, but the overall defect visibility remains limited by background texture and residual noise. In contrast, the fifth principal component (PCA-5th) provides the highest defect contrast and the clearest separation between defective and non-defective regions. In particular, the SISTER-Chirp excitation produces an additional defect indication that is not clearly identifiable in the corresponding linear-scanning and SISTER-Legendre results. This additional feature appears in the highlighted region and becomes distinctly visible only in the PCA-5th reconstruction, suggesting that the chirp-coded illumination improves the excitation of a broader range of thermal diffusion modes and increases the observability of weak subsurface defects. These results demonstrate that the combination of coded SISTER excitation and PCA processing can significantly enhance defect detectability beyond that achievable with conventional linear scanning, with the SISTER-Chirp/PCA-5th combination providing the best overall imaging performance.
\section{Conclusion}
In this work, we proposed structured illumination scanning thermography (SISTER), a photothermal imaging framework that converts conventional temporal modulation into spatial modulation through continuous sample translation. By introducing a structured heating pattern and establishing a Galilean equivalence between spatial scanning and temporal excitation, SISTER unifies scanning thermography and signal-modulated thermography within a common theoretical framework. Based on this formulation, a dynamic-to-static data reconstruction (DSDR) algorithm was developed to transform measurements acquired in the laboratory coordinate system into the material coordinate system, enabling the recovery of transient thermal responses from continuously scanned data.
Numerical simulations and experimental investigations demonstrated that the proposed framework effectively reconstructs stationary thermographic sequences from dynamic scanning measurements. Compared with conventional pulsed thermography, SISTER eliminates stitching artifacts associated with large-area inspection while preserving thermal information comparable to transient excitation methods. The reconstructed temperature histories further confirmed that DSDR converts continuous scanning measurements into pulse-like thermal responses, providing a practical route for large-scale photothermal inspection.
The influence of key system parameters, including scanning velocity and structured illumination width, was systematically investigated. Increasing the scanning velocity was found to reduce thermal contrast because of the shortened thermal interaction time, while simultaneously improving the fidelity of the reconstructed chirp-like excitation waveform. These results reveal a trade-off between signal strength and coding accuracy and provide practical guidelines for parameter optimization. In addition, the analysis of reconstruction errors showed that accurate synchronization between the scanning stage and the reconstruction algorithm is essential for suppressing motion-induced artifacts.
Finally, different spatial coding strategies, including linear, Legendre, and chirp excitations, were implemented within the SISTER framework. The results demonstrate that arbitrary temporal modulation schemes can be realized through equivalent spatial coding patterns without modifying the underlying scanning procedure. In particular, the combination of chirp-coded SISTER and PCA processing provided the highest defect visibility and enabled the detection of additional weak defects that were not clearly observed using conventional linear scanning. These findings highlight the potential of structured spatial coding for enhancing defect detectability and expanding the capabilities of scanning thermographic inspection.

The proposed SISTER framework establishes a general methodology for combining structured illumination, continuous scanning, and thermographic signal processing. Beyond thermography, the theoretical concept of spatially encoded virtual modulation may provide new opportunities for high-speed inspection, large-area nondestructive evaluation, and other diffusion-wave imaging applications.

\begin{acknowledgments}
	This work was supported by the Adolf Martens Fellowship (Grant n. BAM-AMF-2025-1). Parts of this work have been carried out within the project "TTGuss" (IGF 22160 N) funded by the Federal Ministry for Economic Affairs and Energy as part of the "Industrial Collective Research (IGF)" program on the basis of a decision of the German Bundestag.
\end{acknowledgments}


\begin{thebibliography}{99}
	
	\bibitem{1}
	P. Zhu, H. Zhang, S. Sfarra, F. Sarasini, R. Usamentiaga, G. Steenackers, C. Ibarra-Castanedo, X. Maldague,
	\textit{A comprehensive evaluation of the low-velocity impact behaviour of intraply hybrid flax/basalt composites using infrared thermography and terahertz time-domain spectroscopy techniques},
	NDT \& E Int., 154, 103361 (2025).
	
	\bibitem{2}
	P. Zhu, Z. Wei, A. Osman, C. Ibarra-Castanedo, A. Mandelis, X. Maldague, H. Zhang,
	\textit{Real-Time Super-Resolution Imaging System Based on Zero-Shot Learning for Infrared Nondestructive Testing},
	IEEE Tran. Instrum. Meas., 75, 4500409 (2025).
	
	\bibitem{3}
	D. L. Balageas, J. C. Krapez, P. Cielo,
	\textit{Pulsed photothermal modeling of layered material},
	J. Appl. Phys., 59, 348-357 (1986).
	
	\bibitem{4}
	P. Zhu, H. Zhang, S. Sfarra, F. Sarasini, C. Ibarra-Castanedo, X. Maldague, A. Mandelis,
	\textit{Thermal Diffusivity Measurement Based on Thermal / Cooling Excitation: Theory and Experiments},
	IEEE Trans. Instrum. Meas., 2026. https://doi.org/10.1109/TIM.2026.3699727
	
	\bibitem{5}
	P. Zhu, H. Zhang, S. Sfarra, F. Sarasini, R. Usamentiaga, G. Steenackers, C. Ibarra-Castanedo, X. Maldague,
	\textit{Thermal Diffusivity Characterization of Impacted Composites Using Evaporative Cryocooling Excitation and Inverse Physics-Informed Neural Networks},
	IEEE Trans. Instrum. Meas., 75, 6003011 (2026).
	
	\bibitem{6}
	M. Ricci, R. Zito, S. Laureti, 
	\textit{Pseudo-noise pulse-compression thermography: A powerful tool for time-domain thermography analysis},
	NDT \& E Int., 148, 103218 (2024).
	
	\bibitem{7}
	S. Laureti, P. Bison, G. Ferrarini, R. Zito, M. Ricci,
	\textit{Simultaneous multi-frequency lock-in thermography: A new flexible and effective active thermography scheme},
	NDT \& E Int., 146, 103144 (2024).
	
	\bibitem{8}
	Q. Yi, H. Malekmohammadi, G. Y. Tian, S. Laureti, M. Ricci,
	\textit{Quantitative evaluation of crack depths on thin aluminum plate using eddy current pulse-compression thermography},
	IEEE Trans. Ind. Inform., 16, 3963-3973 (2019).
	
	\bibitem{9}
	P. Zhu, R. Wang, K. Sivagurunathan, S. Sfarra, F. Sarasini, C. Ibarra-Castanedo, X. Maldague, H. Zhang, A. Mandelis,
	\textit{Frequency multiplexed photothermal correlation tomography for non-destructive evaluation of manufactured materials},
	Int. J. Extrem. Manuf., 7, 035601 (2025).
	
	\bibitem{10}
	P. Zhu, H. Zhang, S. Sfarra, F. Sarasini, Z. Ding, C. Ibarra-Castanedo, X. Maldague,
	\textit{A novel IR-SRGAN assisted super-resolution evaluation of photothermal coherence tomography for impact damage in toughened thermoplastic CFRP laminates under room and low temperature},
	Compos. Part B-Eng., 316, 113575 (2026).
	
	\bibitem{11}
	P. Zhu, H. Zhang, S. Sfarra, F. Sarasini, R. Usamentiaga, V. Vavilov, C. Ibarra-Castanedo, X. Maldague,
	\textit{Enhancing resistance to low-velocity impact of electrospun-manufactured interlayer-strengthened CFRP by using infrared thermography},
	NDT \& E Int., 144, 103083 (2024).
	
	\bibitem{12}
	P. Burgholzer, L. Gahleitner, G. Mayr,
	\textit{Linkding diffusive fields to virtual waves as their propagative duals},
	Phys. Rev. Applied, 24, 044094 (2025).
	
	\bibitem{13}
	L. Gahleitner, G. Mayr, P. Burgholzer, U. Cakmak,
	\textit{Three-dimensional defect reconstruction in carbon fiber-reinforced composites with temporally non-uniform pulsed thermography data},
	NDT \& E Int., 154, 103363 (2025).
	
	\bibitem{14}
	L. Gahleitner, G. Thummerer, B. Plank, J. Wiedemann, G. Mayr, C. Hühne, P. Burgholzer, U. Cakmak,
	\textit{Photothermal defect imaging in hybrid fiber metal laminates using the vitual wave concept},
	J. Appl. Phys., 135, 074903 (2024).
	
	\bibitem{15}
	D. Zhang, C. Li, C. Zhang, M. N. Slipchenko, G. Eakins, J. X. Cheng,
	\textit{Depth-resolved mid-infrared photothermal imaging of living cells and organisms with submicrometer spatial resolution},
	Sci. Adv., 2 e1600521 (2016).
	
	\bibitem{16}
	G. Jiang, P. Zhu, Y. Gai, et al.,
	\textit{Non-invasive inspection for a hand-bound book of the 19th century: Numerical simulations and experimental analysis of infrared, terahertz, and ultrasonic methods},
	Infrared Phys. Technol., 140, 105353 (2024).
	
	\bibitem{17}
	G. Jiang, P. Zhu, S. Sfarra, et al.,
	\textit{Faster R-CNN-CA and thermophysical properties of materials: An ancient marquetry inspection based on infrared and terahertz techniques},
	Infrared Phys. Technol., 142, 105563 (2024).
	
	\bibitem{18}
	L. Cheng, M. Kersemans,
	\textit{Dual-IRT-GAN: A defect-aware deep adversarial network to perform super-resolution tasks in infrared thermographic inspection},
	Compos. Part B-Eng., 247, 110309 (2022).
	
	\bibitem{19}
	P. Zhu, H. Zhang, S. Sfarra, F. Sarasini, X. Maldague, A. Mandelis,
	\textit{Three-dimensional wide-bandwidth quantum energy truncation terahertz coherence tomography},
	Phys. Rev. Lett., (2026). https://doi.org/10.1103/spsr-xr47
	
	\bibitem{20}
	C. Holmes, B. W. Drinkwater, P. D. Wilcox,
	\textit{Post-processing of the full matrix of ultrasonic transmit-receive array data for non-destructive evaluation},
	NDT \& E Int., 38, 701-711 (2005).
	
	\bibitem{21}
	P. Zhu, Z. Wei, S. Sfarra, R. Usamentiaga, G. Steenackers, A. Mandelis, X. Maldague, H. Zhang,
	\textit{THz-Super-Resolution Generative Adversarial Network: Deep-Learning-Based Super-Resolution Imaging Using Terahertz Time-Domain Spectroscopy},
	IEEE Trans. Ind. Inform., 21, 6660-6669 (2025).
	
	\bibitem{22}
	X. Maldague, S. Marinetti,
	\textit{Pulse phase infrared thermography},
	J. Appl. Phys., 79, 2694-2698 (1996).
	
	\bibitem{23}
	J. Lecompagnon, S. Ahmadi, P. Hirsch, C. Rupprecht, M. Ziegler,
	\textit{Thermographic detection of internal defects using 2D photothermal super resolution reconstruction with sequential laser heating},
	J. Appl. Phys., 131, 185107 (2022).
	
	\bibitem{24}
	N. Tabatabaei, A. Mandelis,
	\textit{Thermal Coherence Tomography Using Match Filter Binary Phase Coded Diffusion Waves},
	Phys. Rev. Lett., 107, 165901 (2011).
	
	\bibitem{25}
	A. Mandelis, L. Nicolaides, Y. Chen,
	\textit{Structure and the Reflectionless/Refractionless Nature of Parabolic Diffusion-Wave Fields},
	Phys. Rev. Lett., 87, 020801 (2001).
	
	\bibitem{26}
	C. Ibarra-Castanedo, J. M. Piau, S. Guilbert, N. P. Avdelidis, M. Genest, A. Bendada, X. P. V. Maldague,
	\textit{Comparative study of active thermography techniques for the nondestructive evaluation of honeycomb structures},
	Res. Nondestruct. Eval., 20, 1-31 (2009).
	
	\bibitem{27}
	C. Ibarra-Castanedo, D. González, M. Klein, S. Vallerand, X. Maldague,
	\textit{Infrared image processing and data analysis},
	Infrared Phys. Technol., 46, 75-83 (2004).
	
	\bibitem{28}
	A. Mandelis, D. Thapa,
	\textit{Generalized Fourier-Laplace photothermal spectroscopy of optically absorbing media generated by arbitrary optical-excitation waveforms},
	Phys. Rev. Applied, 23, 054034 (2025).
	
	\bibitem{29}
	A. Mandelis, D. Thapa,
	\textit{Generalized Fourier-Laplace photothermal spectroscopy of optically absorbing media generated by arbitrary optical-excitation waveforms},
	Phys. Rev. Applied, 23, 054034 (2025).
	
	\bibitem{30}
	D. Thapa, P. Tavakolian, G. Zhou, et al.,
	\textit{Three-dimensional thermophotonic super-resolution imaging by spatiotemporal diffusion reversal method},
	Sci. Adv., 9, eadi1899 (2023).
	
	\bibitem{31}
	S. Ahmadi, G. Thummerer, S. Breitwieser, et al.,
	\textit{Multidimensional Reconstruction of Internal Defects in Additively Manufactured Steel Using Photothermal Super Resolution Combined With Virtual Wave-Based Image Processing},
	IEEE Trans. Ind. Inform., 17, 7368-7378 (2021).
	
	\bibitem{32}
	L. M. Wilcox, M. Bonmarin, K. M. Donnell,
	\textit{Effect of Signal Modulation on Active Microwave Thermography},
	IEEE Trans. Instrum. Meas., 73, 7005314 (2024).
	
	\bibitem{33}
	K. Tian, J. Peng, Q. Zhang, F. Zhang, J. Lee,
	\textit{Laser arrays scanning thermography with optimized excitation signals for efficient rail defect detection},
	NDT \& E Int., 156, 103461 (2025).
	
	\bibitem{34}
	S. Schmid, J. Reinhardt, C. U. Grosse,
	\textit{Spatial and temporal deep learning for defect detection with lock-in thermography},
	NDT \& E Int., 143, 103063 (2024).
	
	\bibitem{35}
	P. Zhu, D. Wu, Y. Wang, Z. Miao,
	\textit{Defect detectability based on square wave lock-in thermography},
	Appl. Opt., 61, 6134-6143 (2022).
	
	\bibitem{36}
	S. Hedayatrasa, G. Poelman, J. Segers, W. V. Paopegem, M. Kersemans,
	\textit{Novel discrete frequency-phase modulated excitation waveform for enhanced depth resolvability of thermal wave radar},
	Mech. Syst. Signal Process., 132, 512-522 (2019).
	
	\bibitem{37}
	N. Tabatabaei, A. Mandelis,
	\textit{Thermal-wave radar: A novel subsurface imaging modality with extended depth-resolution dynamic range},
	Rev. Sci. Instrum., 80, 034902 (2009).
	
	\bibitem{38}
	F. Wang, Y. Wang, J. Liu, Y. Wang,
	\textit{The feature recognition of CFRP subsurface defects using low-energy chirp-pulsed radar thermography},
	IEEE Trans. Ind. Inform., 16, 5160-5168 (2019).
	
	\bibitem{39}
	S. Hedayatrasa, W. V. Paepegem, M. Kersemans,
	\textit{Diffusion-compensated correlation analysis of frequency-modulated thermal signal for quantitative infrared thermography},
	Mech. Syst. Signal Process., 197, 110373 (2023).
	
	\bibitem{40}
	C. Ibarra-Castanedo, P. Servais, M. Klein, T. Boulanger, A. Kinard, S. Hoffait, X. P. V. Maldague,
	\textit{Detection and Characterization of Artificial Porosity and Impact Damage in Aerospace Carbon Fiber Composites by Pulsed and Line Scan Thermography},
	Appl. Sci., 13, 6135 (2023).
	
	\bibitem{41}
	R. Usamentiaga, C. Ibarra-Castanedo, X. Maldague, 	
	\textit{More than Fifty Shades of Grey: Quantitative Characterization of Defects and Interpretation Using SNR and CNR}, 	
	NDT \& E Int., 145, 103133 (2024).
	
	\bibitem{42}
	N. Rajic, 	
	\textit{Principal component thermography for flaw contrast enhancement and flaw depth characterisation in composite structures}, 	
	Compos. Struct., 58, 521-528 (2002).

	
\end{thebibliography}
\end{document}